\documentclass[twoside,a4paper,11pt]{sea10}
\usepackage{graphicx}
\usepackage{hyperref}
\usepackage{movie15}
\newcommand\ageL{$\langle {\rm log}\,age\rangle _L$}

\newcommand\HLR{$a_{50}^L$}

\newcommand\logZM{$\langle  \log Z_\star \rangle _M$}

\begin{document}
\pagenumbering{arabic}
\pagestyle{myheadings}
\thispagestyle{empty}
{\flushleft\includegraphics[width=\textwidth,bb=58 650 590 680]{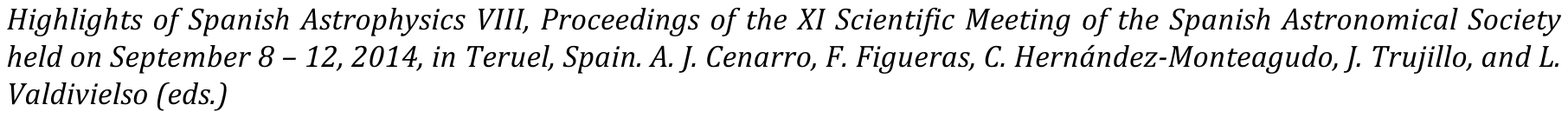}}
\vspace*{0.2cm}
\begin{flushleft}
{\bf {\LARGE
%
CALIFA across the Hubble types: Spatially resolved properties of the stellar populations
%
}\\
\vspace*{1cm}
%
Gonz\'alez Delgado, R.M.$^{1}$,
Garc\'\i a-Benito, R.$^{1}$, 
P\'erez, E.$^{1}$,
Cid Fernandes, R.$^{2}$,
de Amorim, A.L.$^{2}$,
Cortijo-Ferrero, C.$^1$,
Lacerda, E.A.D.$^{1,2}$,
L\'opez Fern\'andez, R.$^1$,
S\'anchez, S.F.$^{3}$,
Vale Asari, N.$^2$,
and
CALIFA collaboration
%
}\\
\vspace*{0.5cm}
%
$^{1}$
Instituto de Astrof\'{\i}sica de Andaluc\'{\i}a (CSIC), P.O. Box 3004, 18080 Granada, Spain\\
$^{2}$
Departamento de F\'{\i}sica, Universidade Federal de Santa Catarina, P.O. Box 476, 88040-900, Florian\'opolis, SC, Brazil\\
$^{3}$Instituto de Astronom\'\i a, Universidad Nacional Auton\'oma
de M\'exico, A.P. 70-264, 04510, M\'exico,D.F.
%
\end{flushleft}
%
\markboth{
CALIFA across the Hubble type
}{ 
%
Gonz\'ales Delgado et al. 
%
}
\thispagestyle{empty}
\vspace*{0.4cm}
\begin{minipage}[l]{0.09\textwidth}
\ 
\end{minipage}
\begin{minipage}[r]{0.9\textwidth}
\vspace{1cm}
\section*{Abstract}{\small
%
We analyze the spatially resolved star formation history of 300 nearby galaxies from the CALIFA integral field spectroscopic survey to investigate the radial structure and gradients of the present day stellar populations properties as a function of Hubble type and galaxy stellar mass. A fossil record method based on spectral synthesis techniques is used to recover spatially and temporally resolved maps of stellar population properties of spheroidal and spiral galaxies with masses $10^9$ to $7 \times 10^{11}$ M$_\odot$. The results show that galaxy-wide spatially averaged stellar population properties (stellar mass, mass surface density, age, metallicity, and  extinction) match those obtained from the integrated spectrum, and that these spatially averaged properties match those at $R = 1$ HLR (half light radius), proving that the effective radii are really effective.
 Further, the individual radial profiles of the stellar mass surface density ($\mu_\star$), luminosity weighted ages (\ageL), and mass weighted metallicity (\logZM) are stacked in bins of  galaxy morphology (E, S0, Sa, Sb, Sbc, Sc and Sd). All these properties show negative gradients as a sign of the inside-out growth of massive galaxies. However, the gradients depend on the Hubble type in different ways. For the same galaxy mass, E and S0 galaxies show the largest inner gradients in $\mu_\star$; while MW-like galaxies (Sb with $M_\star \sim 10^{11} M_\odot$)  show the largest inner age and metallicity gradients. The age and metallicity gradients suggest that major mergers have a relevant role in growing the center (within 3 HLR) of massive early type galaxies; and radial mixing may play a role flattening the radial metallicity gradient in MW-like disks.
%
\normalsize}
\end{minipage}
%
%
%
\section{Introduction \label{intro}}
Galaxies are a complex mix of stars, interstellar gas, dust, and dark matter, distributed in different components (bulge, disk, and halo) whose present day structure and dynamics  are intimately linked to the assembly and evolution of galaxies over the age of the Universe. Different observational and theoretical approaches can be followed to know how galaxies form and evolve. Spatially resolved and cosmic time information of the properties of galaxies in the local universe can provide very useful observational constraints for the galaxy formation models.

One important step forward can be achieved with the use of Integral Field Spectroscopy (IFS), that provides a unique 3D view of galaxies (two spatial plus one spectral dimensions). Using the fossil record of the stellar populations imprinted in their spectra,  galaxies can be dissected in space and time providing 3D information to retrieve when and how the mass and stellar metallicity were assembled as a function of look-back time.

Until very recently it was not possible to obtain IFS data for samples larger than a few tens of galaxies. The ATLAS3D \cite{Cappellari11}, CALIFA \cite{Sanchez12}, SAMI \cite{Croom14} and MaNGA \cite{Bundy14} surveys are the first in doing this step forward, gathering data for hundreds to thousands of galaxies in the nearby Universe.

CALIFA, observing 600 nearby galaxies with the PPaK IFU, is our currently-ongoing survey, designed to make a rational and efficient use of the singular and unique facility provided by the 3.5m telescope at the Calar Alto observatory. 
 CALIFA is the first spectroscopic survey that can obtain global and spatially resolved properties of galaxies. These features make CALIFA unique for studies of galaxy formation and evolution.

\section{Current status of CALIFA} 

CALIFA has been designed to allow the building of 2D maps of: a) stellar populations and their time evolution; b) ionized gas distribution, excitation and abundances; c) stellar and ionized gas kinematics. The survey obtains spatially resolved spectroscopy of a diameter selected sample of 600 galaxies in the local universe (0.005$\leq z \leq$ 0.03). CALIFA uses the PPak IFU, with an hexagonal FoV$\sim$1.3 sq.arcmin, with a 100$\%$ covering (using a three-pointing dithering scheme). The  wavelength range covered is 3700$-$7000 \AA, with two overlapping setups at  resolutions $R\sim850$ and $R\sim1650$, with FWHM $\sim$6 \AA\ and 2.3 \AA, respectively. 
 
CALIFA is a legacy project, intended for the community. The reduced data is released once the quality is guaranteed. The first Data Release (DR1) occurred in October 2012 (http://califa.caha.es/DR1) in a timely manner. DR2, released in 2014 October 1st, delivers 400 fully flux calibrated (with the new pipeline 1.5) data-cubes of 200 galaxies \cite{RGB14}.
 Extended descriptions of the survey, data reduction and sample can be found in \cite{Sanchez12}, \cite{Husemann13}, \cite{RGB14} and \cite{Walcher14}.
 
 \section{Stellar populations analysis: method and previous results}
 
We use the {\sc starlight} code \cite{Cid05} to do a  $\lambda$-by-$\lambda$ spectral fit using different sets of single stellar population (SSP) models. These SSPs are from a combination of \cite{Vazdekis10} and \cite{RGD05} (labelled {\it GMe}), or from \cite{Charlot07} (labelled {\it CBe}).

Our scientific results from the first 100-300 CALIFA galaxies were presented in \cite{Perez13}, \cite{Cid13,Cid14} and \cite{RGD14a,RGD14b}. The spatially resolved history of the stellar mass assembly is obtained for galaxies beyond the Local Group. Highlight results of these works are:
1) Massive galaxies grow their stellar mass inside-out. The negative radial gradients of the stellar population ages also supports this scenario.
2) For the first time, the signal of downsizing is shown to be spatially preserved, with both inner and outer regions growing faster for more massive galaxies.
3) The relative growth rate of the spheroidal component (nucleus and inner galaxy), which peaked 5--7 Gyr ago, shows a maximum at a critical stellar mass $\sim 7 \times 10^{10}  M_\odot$. 
4) Galaxies are about 15$\%$ more compact in mass than in light; for disk galaxies this fraction increases with galaxy mass.
5) A systematic study of the uncertainties associated with the spectral synthesis applied to data cubes.
6) Global (and local) stellar mass (mass surface density) metallicity relations and their evolution are derived.
7) In disks, the stellar mass surface density regulates  ages and  stellar metallicity. In spheroids, the total stellar mass dominates the physics of star formation and chemical enrichment.
8) In terms of integrated versus spatially resolved properties, the stellar population properties are well represented by their values at 1 half light radius (HLR). 
Fig.\ 1 updates this particular result. With 300 galaxies, the plots confirm that the galaxy-wide spatially averaged stellar population properties (stellar galaxy mass, stellar mass surface density, age, stellar metallicity, and stellar extinction) match those obtained from the integrated spectrum as well as the values  at 1 HLR, proving that  effective radii are really effective.



\begin{figure*}
\includegraphics[width=\textwidth]{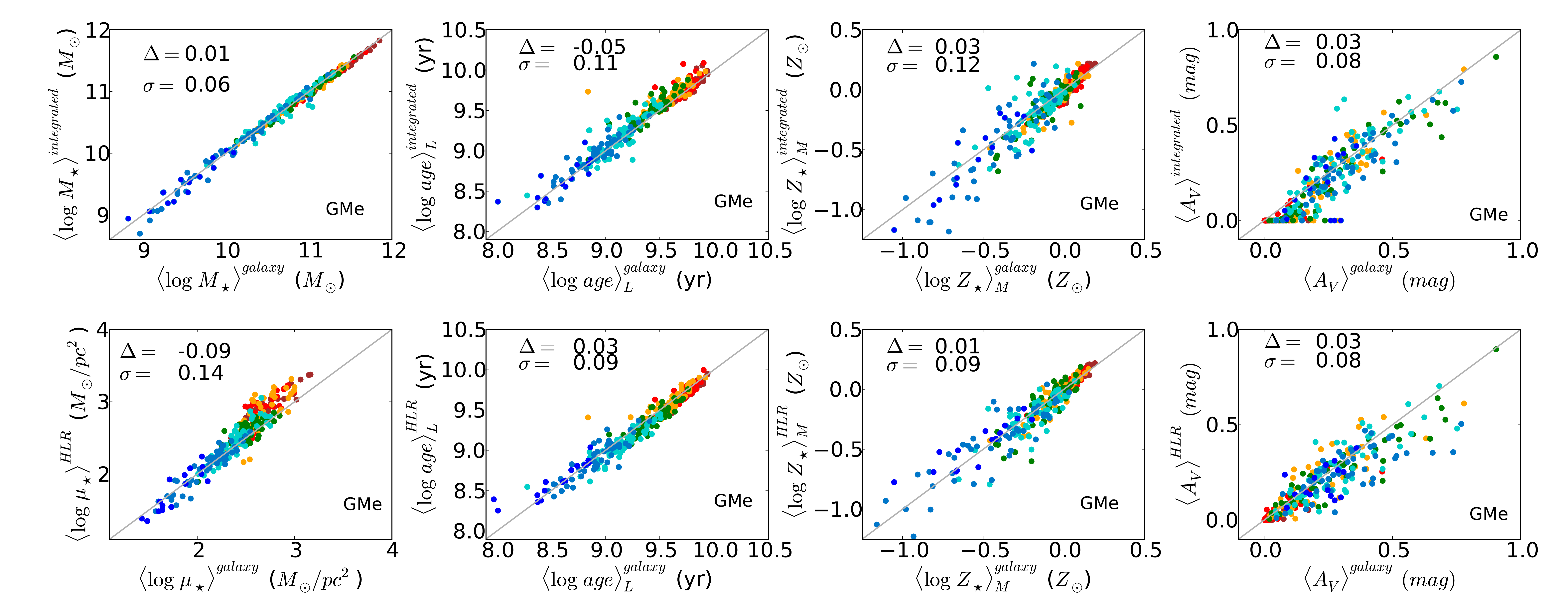}
\caption{Comparison of the galaxy-wide average stellar population properties derived from the spatially resolved spectral synthesis analysis and the properties obtained at 1 HLR (lower panel) or from the integrated spectra (upper panel). From left to right are the galaxy  mass (upper panel), or stellar mass surface density (lower panel), age, metallicity, and extinction.}
\label{fig:integrated}
\end{figure*}

\section{Stellar populations properties across the Hubble sequence}

A  step to understand how galaxies form and evolve is classifying them and studying  their properties. Most of the massive galaxies in the near Universe are E, S0 and spirals \cite{Blanton09}, following well the Hubble tuning-fork diagram.  The bulge fraction seems to be one of the main physical parameters that produce the Hubble sequence, increasing from late to early spirals. In this scheme, S0 galaxies are a transition class between  spirals  and  ellipticals, with large bulges, but intermediate between Sa and E galaxies. On the other hand,  properties such as color, mass, surface brightness, luminosity, and gas fraction are correlated with the Hubble type \cite{Roberts94}. This suggests that the Hubble sequence can illustrate possible paths for galaxy formation and evolution. If so, how are  the spatially resolved stellar population properties of galaxies correlated with the Hubble type? Can the Hubble-tuning-fork scheme be useful to organize galaxies by galaxy mass and age or  galaxy mass and metallicity? 

CALIFA is a suitable laboratory to address these questions because it includes a significant amount of E, S0 and spirals. After a visual classification, the 300 galaxies were grouped in 41 E,  32 S0, 51 Sa, 53 Sb, 58 Sbc, 50 Sc, and 15 Sd. This sub-sample is representative of the morphological  distribution of the whole CALIFA sample (Fig.\ 2a). In terms of galaxy stellar mass, $M_\star$, this sub-sample ranges from 10$^9$ to 7$\times$10$^{11}$ M$_\odot$ (for {\it GMe} SSPs) (Fig.\ 2b). There is a clear segregation in mass: galaxies with high bulge-to-disk ratio (E, S0, Sa) are the most massive ones ($\geq$ 10$^{11}$ M$_\odot$), and galaxies with small bulges (Sc-Sd) have masses  $M \leq 10^{10}  M_\odot$. The stellar mass distribution obtained with {\it CBe} models is similar to that obtained  with the {\it GMe} base, but shifted by $-0.25$ dex due to the difference in IMF (Chabrier in {\it CBe} versus Salpeter in {\it GMe}).

Below we present the radial structure of the stellar mass surface density ($\mu_\star$), luminosity weighted stellar age (\ageL),  and mass weighted stellar metallicity (\logZM),  stacking the galaxies by their Hubble type.  Most of the results discussed here are obtained with the  {\it GMe} SSP models, but similar results are obtained with the {\it CBe} base (see Fig.\ 2b).

\begin{figure*}
\includegraphics[width=0.48\textwidth]{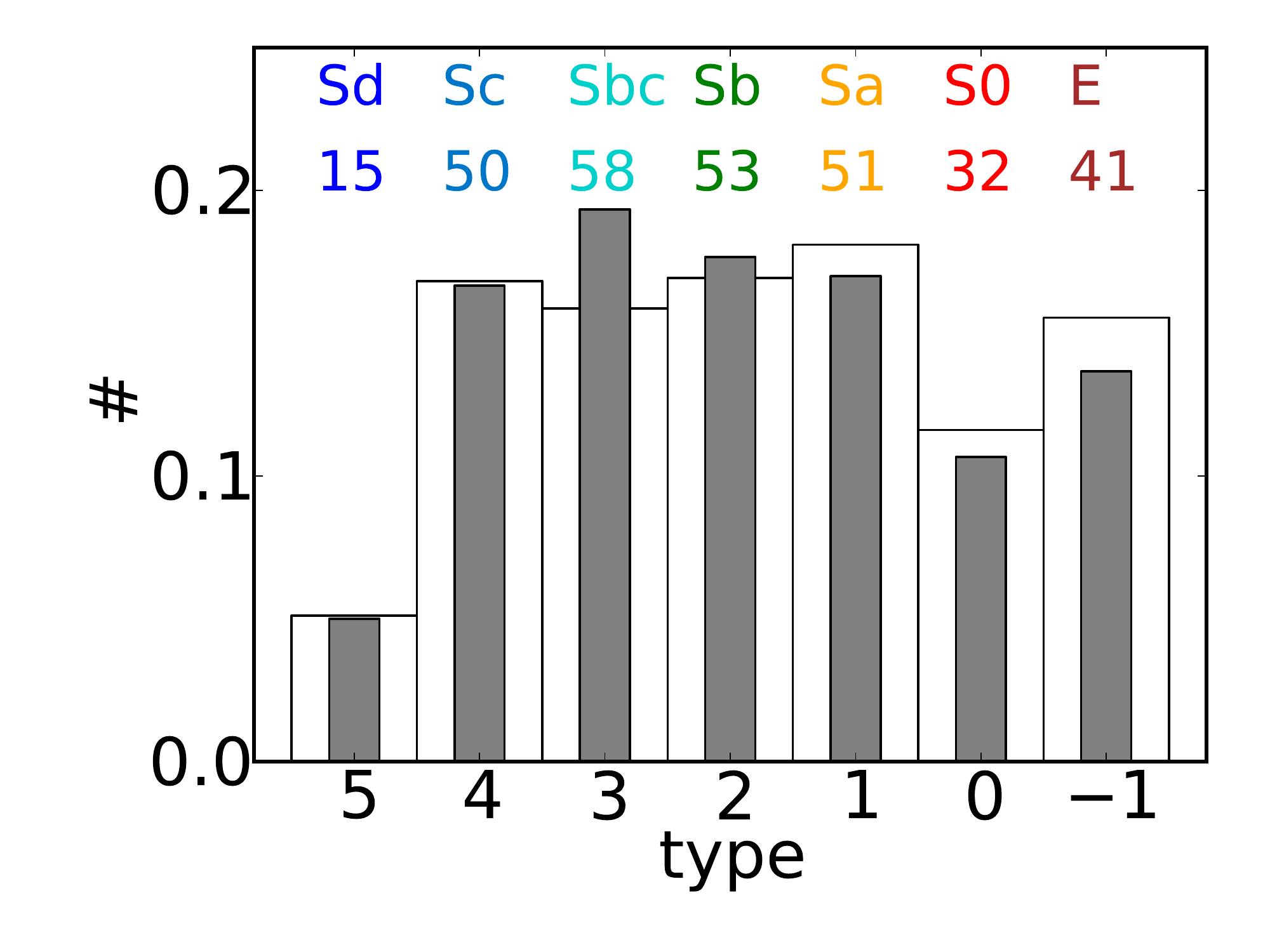}
\includegraphics[width=0.48\textwidth]{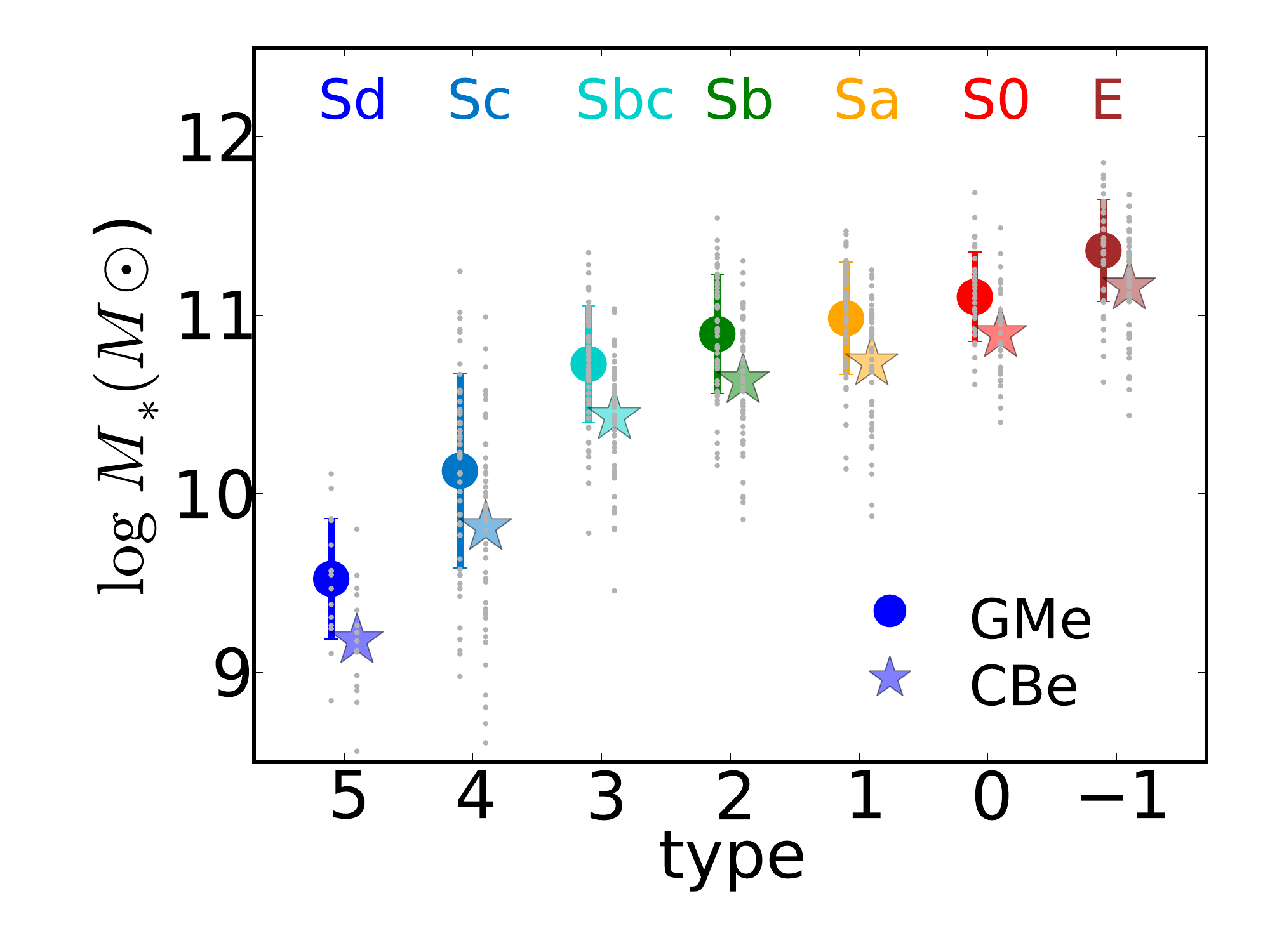}
\caption{a)  Distribution of Hubble types in the CALIFA mother sample (empty bars) and the 300 galaxies analyzed here (filled bars). The number of galaxies in our sample are labeled in colors.  b) Distribution of the galaxy stellar masses obtained from the spatially resolved spectral fits of each galaxy for each Hubble type of the galaxies in this work (grey small points). The colored dots (stars) are the mean galaxy stellar mass in each Hubble type obtained with the {\it GMe} ({\it CBe}) SSP models. The bars show the dispersion in mass. }
\label{fig:histmass}
\end{figure*}

\subsection{Stellar mass surface density}

Fig.\ 3a shows the radial profiles (in units of HLR, a$_{50}^L$) of $\log\ \mu_\star$. Individual results are stacked in seven morphological bins. The error bar indicate the dispersion at $R = 1$ HLR distance in Sa galaxies, but it is similar for other Hubble types and radial distances.  Negative gradients are detected in all galaxy types and steepen from late type (Sd) to early type (S0, and E) galaxies. At a constant M$_\star$, spheroidals (S0 and E) are more compact than spirals;  S0 and E  have  similar compactness at all distances (Fig.\ 3b). 

\begin{figure*}
\includegraphics[width=0.48\textwidth]{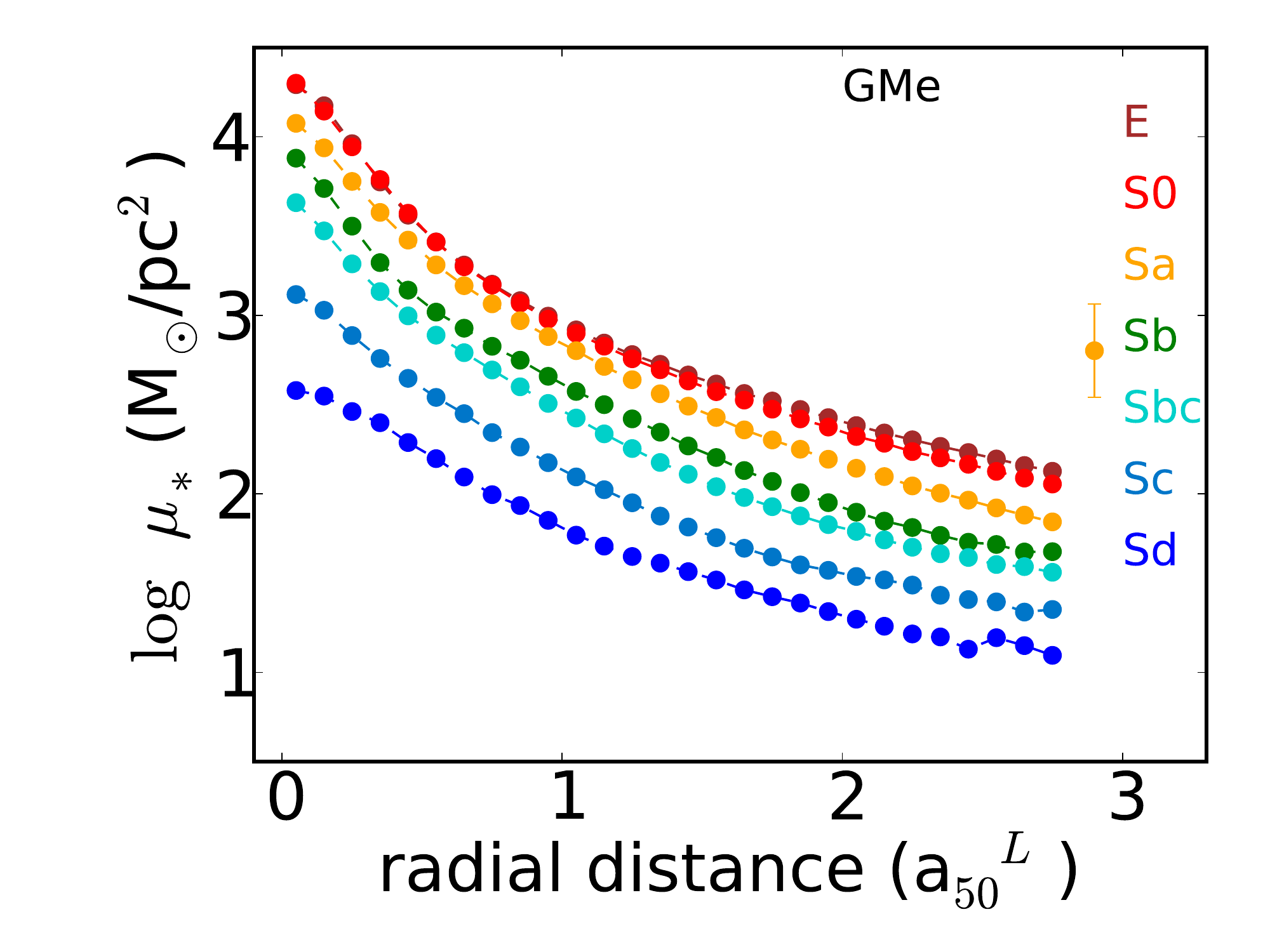}
\includegraphics[width=0.48\textwidth]{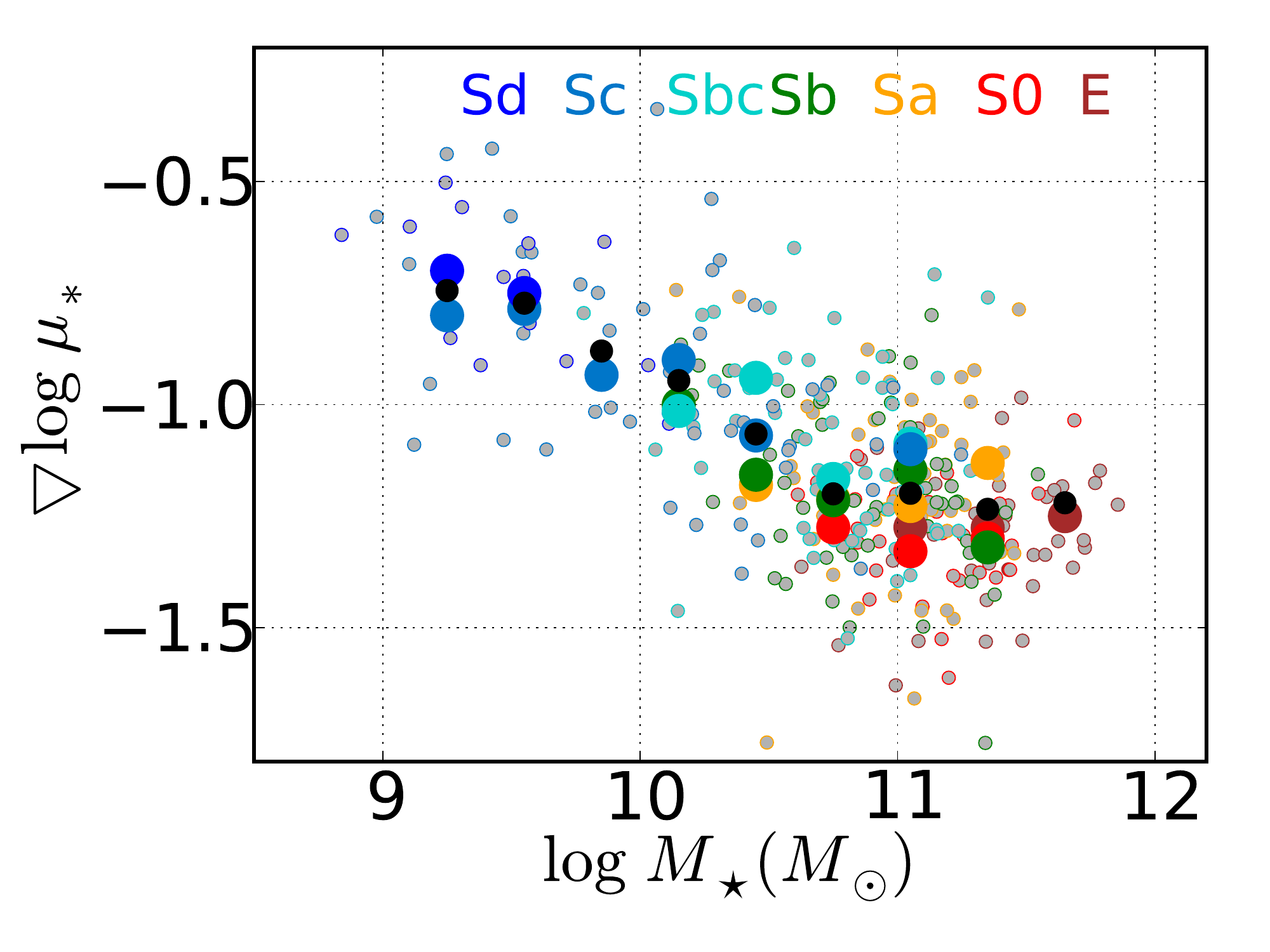}
\caption{Left: The radial profiles (in units of HLR) of the stellar mass surface density stacking in seven morphological type bins. The error bar indicates the dispersion at 1 HLR in  Sa galaxies. Right: Correlation between the inner gradient of $\log\ \mu_\star$ and the galaxy stellar mass. Small circles represent the results for each galaxy. Large circles represent the averaged inner gradient in mass intervals of 0.25 dex for each color-coded morphological type. Grey dots show the averaged correlation between the inner gradient of $\log \mu_\star$ and galaxy mass independently of the morphological type.}
\label{fig:integrated}
\end{figure*}

\subsection{Stellar ages}

Fig.\ 4a shows the radial profiles of  \ageL. Symbols are as in Fig.\ 3a. Negative gradients are detected for all  Hubble types, suggesting that  quenching progresses outwards, and  galaxies  grow inside-out, as we concluded with  our mass assembly growth analysis \cite{Perez13}. Inner gradients are calculated between the galaxy nucleus and  1 \HLR, and the outer gradient between 1 and 2 \HLR. The inner age gradient shows a clear behaviour with Hubble type, with maximum for intermediate  type spirals (Sb--Sbc). At constant $M_\star$, Sb--Sbc galaxies have the largest age gradient (Fig.\ 4b). The age gradient in the outer disk (between $R = 1$ and 2 \HLR) is smaller than the inner ones, but again it is largest amongst Sa-Sb-Sbc's. 

\begin{figure*}
\includegraphics[width=0.48\textwidth]{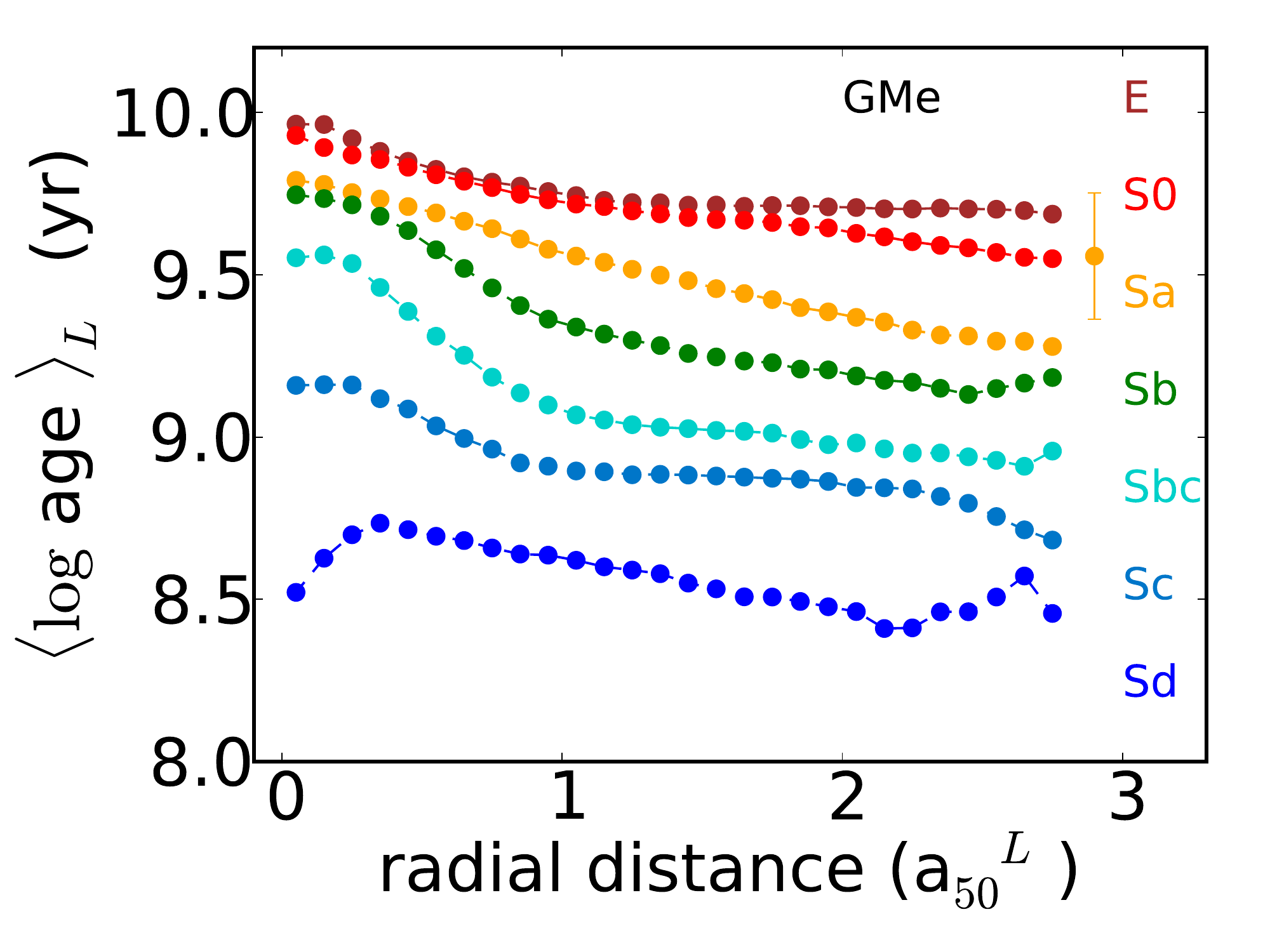}
\includegraphics[width=0.48\textwidth]{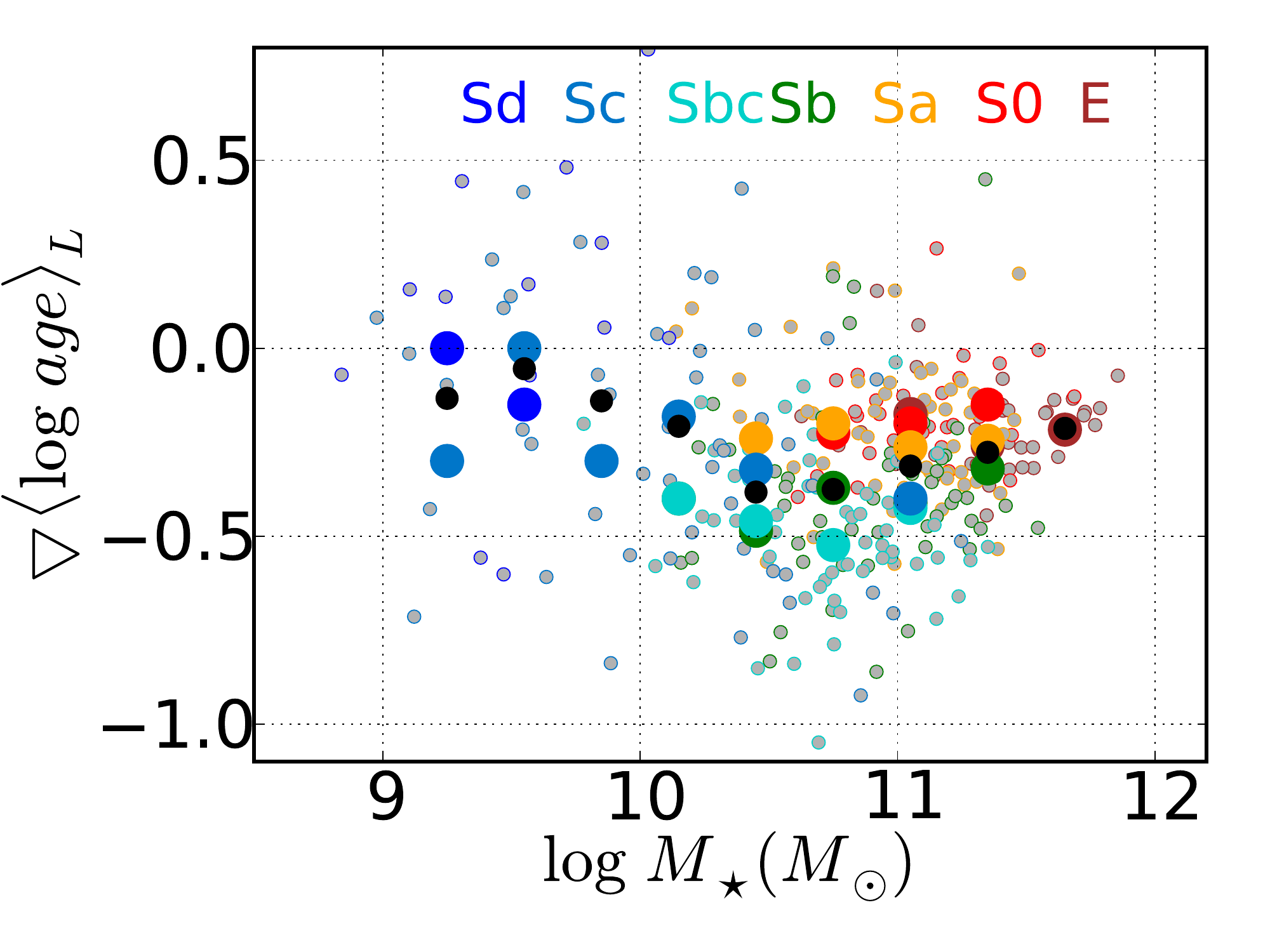}
\caption{As Fig.\ 3 but for (luminosity weighed) age  of the stellar population.  }
\label{fig:integrated}
\end{figure*}

\subsection{Stellar metallicity}
Fig.\ 5a shows the radial profiles of mass weighted stellar metallicity obtained as explained in \cite{RGD14b}. Except for late types (Sc--Sd), spirals present negative $Z_\star$ gradients: On  average $\sim -0.1$ dex per \HLR, similar to the value obtained for the nebular oxygen abundances obtained by CALIFA \cite{Sanchez13}. For galaxies of equal stellar mass, the  intermediate type spirals (Sbc) have the largest gradients. These negative gradients again are a sign of the inside-out growth of the disks.

\begin{figure*}
\includegraphics[width=0.48\textwidth]{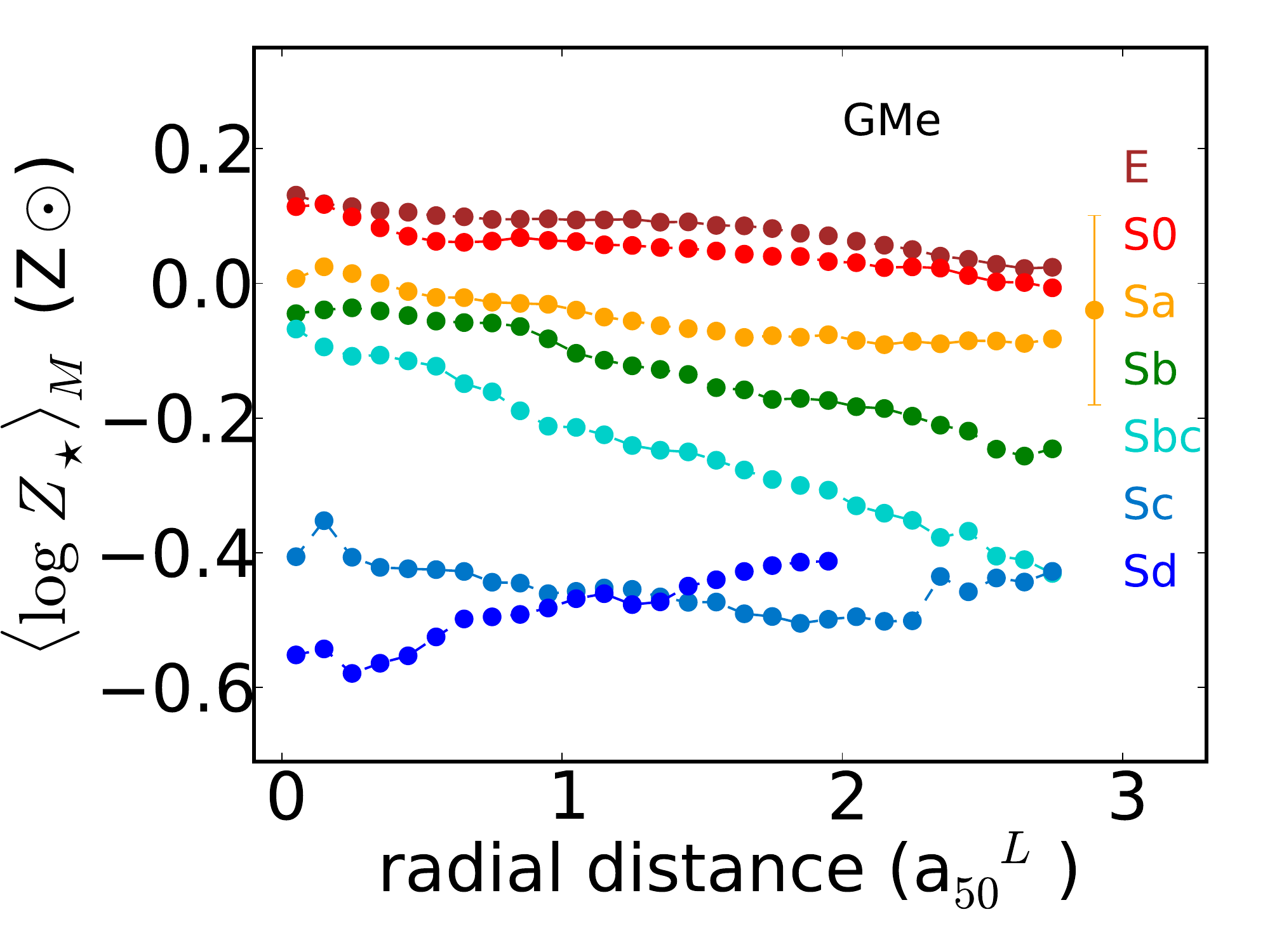}
\includegraphics[width=0.48\textwidth]{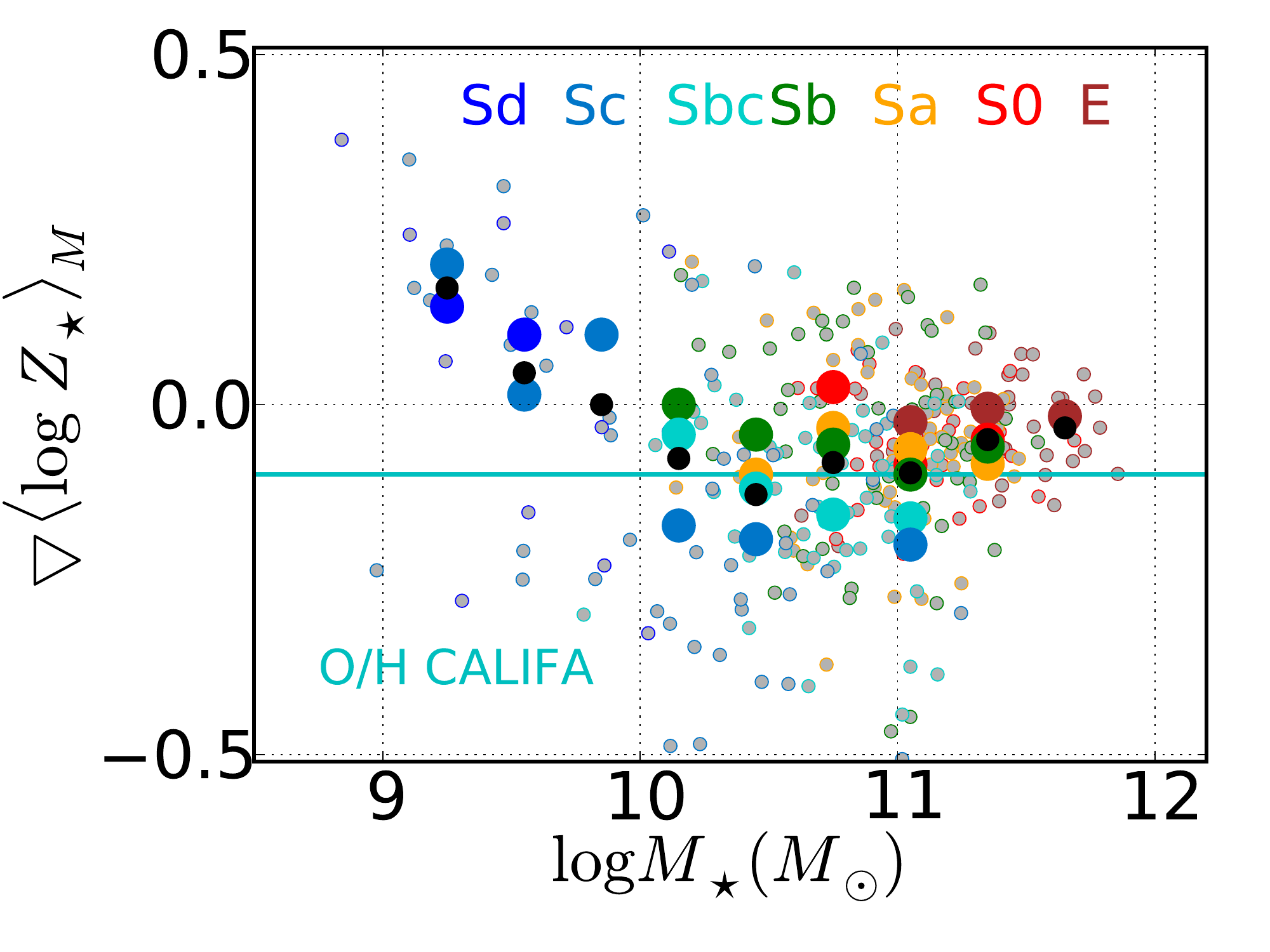}
\caption{As Fig.\ 3 but for the (mass weighed)  stellar metallicity.   }
\label{fig:integrated}
\end{figure*}

\section{Discussion}

Models predict that massive early type galaxies form in a two-phase scenario. First, galaxies assembled their mass through dissipative processes and in-situ star formation. The star formation is induced by cold flow accretion or by gas-rich mergers. Then, ex-situ star formation formed by  minor mergers of galaxies with the central more massive one. Because the central core of these ETG have enriched very quickly due to major mergers that happened at higher redshift, and the accreted satellites have fewer metals than the central massive one but they have old stellar populations, we expect negative metallicity radial gradients, and positive age radial gradients towards the outskirts of the galaxy.

Ages and metallicity radial profiles of E and S0  are negative but shallower than expected if minor mergers are relevant in growing the central 1-3 HLR of these massive early type galaxies. Our results indicate that $\Delta \langle  \log Z_\star \rangle _M / \Delta log R \ge -0.1$,  more in agreement with  the SPH chemodynamical simulations by \cite{Kobayashi04} if major mergers are relevant in the formation of E-SO galaxies. 


Like massive ETGs, spiral galaxies also formed in two phases. In the first phase the bulge formed in a similar way as the core of S0 and E. In the second phase, the disk grows by star formation in-situ from  infalling gas. Metal poor gas  with higher angular momentum at lower redshifts is turned into stars at larger  radii. Negative radial ages and metallicity gradients are expected. This is in very good agreement with our results (see Fig.\ 4 and Fig.\ 5). However, our Sb-Sbc galaxies have metallicity gradients that are somewhat flatter ($-0.025$  dex/kpc) than the prediction by the classical chemical evolution models  \cite{Chiappini01,Molla05}. However, our results go in line with the recent N-body hydrodynamical simulations by \cite{Minchev14}, built to explain the variation of the MW metallicity radial gradient as a function of the height above the Galactic plane. Radial mixing due to stellar migration may also be relevant in flattening the radial metallicity gradients in disks \cite{Roskar08}. A future study of the time evolution of the spatially resolved stellar metallicity in CALIFA spiral galaxies will help to understand if the present metallicity gradient results from a flat gradient that steepens as a consequence of stellar mixing; or it steepens as a consequence of very efficient feedback processes in the star formation processes.

%
%
\small  
%
\section*{Acknowledgments}   
%
CALIFA is the first legacy survey carried out at Calar Alto. The CALIFA collaboration would like to thank the IAA-CSIC and MPIA-MPG as major partners of the observatory, and CAHA itself, for the unique access to telescope time and support in manpower and infrastructures.  We also thank the CAHA staff for the dedication to this project.
Support from the Spanish Ministerio de Econom\'\i a y Competitividad, through projects AYA2010-15081 (P.I. RGD), AYA2010-22111-C03-03, and AYA2010-10904E (SFS). RGD acknowledges the support of CNPq (Brazil) through Programa Ciencia sem Fronteiras (401452/2012-3).
%

%
\end{document}